\documentclass[twocolumn]{emulateapj}

\usepackage{epsfig}
\usepackage{epstopdf}
\usepackage{natbib}
\usepackage{float}
\usepackage[usenames]{color}
\usepackage{color,epsfig,amsmath,amssymb,fancyhdr}  
\usepackage{ulem}  
\usepackage[squaren, Gray, cdot]{SIunits}
\usepackage[fixamsmath]{mathtools}
\newcommand{\beq}{\begin{equation}}
\newcommand{\eeq}{\end{equation}}
\newcommand{\ben}{\begin{eqnarray}}
\newcommand{\een}{\end{eqnarray}}
\newcommand{\bi}{\begin{itemize}}
\newcommand{\ei}{\end{itemize}}
\newcommand{\galp}{photon/ALP }

\shorttitle{ALP constraints from Hydra A}
\shortauthors{Wouters \& Brun}

\begin{document}

\title{Constraints on axion-like particles from X-ray observations of the Hydra galaxy cluster}

\author{Denis Wouters\altaffilmark{1} and Pierre Brun\altaffilmark{2}}
\affil{CEA, Irfu, Centre de Saclay, F-91191
Gif-sur-Yvette | France}
 
\altaffiltext{1}{denis.wouters@cea.fr}
\altaffiltext{2}{pierre.brun@cea.fr}

\begin{abstract}
Axion-like particles (ALPs) belong to a class of new pseudoscalar particles that generically couple to photons, opening the possibility of oscillations from photons into ALPs in an external magnetic field. Having witnessed the turbulence of their magnetic fields, these oscillations are expected to imprint irregularities in a limited energy range of the spectrum of astrophysical sources. In this study, \textit{Chandra} observations of the Hydra galaxy cluster are used to constrain the value of the coupling of ALPs to photons. We consider the conversion of X-ray photons from the central source Hydra A in the magnetic field of the cluster.  The magnetic field strength and structure are well determined observationally, which adds to the robustness of the analysis. The absence of anomalous irregularities in the X-ray spectrum of Hydra A conservatively provides the most competitive constraints on the coupling constant for ALP masses below $7\times 10^{-12}\;\rm eV$ at the level of $g_{\gamma a} < 8.3 \times 10^{-12}\;\rm GeV^{-1}$ at the 95\% confidence level. Because of the specific phenomenology involved, these constraints actually hold more generally for very light pseudo-Nambu-Goldstone bosons.
\end{abstract}

\maketitle

\section{Introduction}
\label{sec:intro}

Many extensions of the standard model of particle physics predict the existence of new light bosons. One of the most studied examples is the axion, a pseudoscalar particle associated with the spontaneous breaking of the U(1) Peccei-Quinn symmetry \citep{Peccei:1977hh}. The symmetry was introduced as a solution to the strong CP problem, with the spontaneous breaking of the symmetry dynamically tuning the CP angle to zero \citep{Peccei:1977ur}. In this case, the mass of the axion is predicted to scale with the inverse symmetry breaking scale $f$. Pseudoscalar particles generically couple with the electromagnetic field via a two photon vertex of the form:
\beq
\mathcal{L}_{\gamma a} = -\frac{1}{4}g_{\gamma a}F_{\mu\nu}\tilde{F}^{\mu\nu} a = g_{\gamma a} \vec{E}\cdot \vec{B} a \;\;,
\label{eq:lagrangian}
\eeq
where $F$ is the electromagnetic tensor, $\tilde{F}$ is its dual, $\vec{E}$ and $\vec{B}$ are the electric and magnetic fields, $a$ is the axion field, and $g_{\gamma a}$ is the axion-photon coupling  strength. For standard axions, $g_{\gamma a}$  also scales with $1/f$ so that the coupling strength and the mass of the hypothetical particle are proportional. However pseudoscalar particles with unrelated mass and coupling strength can be expected from more general U(1) symmetries \citep{Kim:1986ax, 2012PDU.....1..116R} or in extra-dimension gauge theories \citep{1996PhRvL..76.1015T}. Those have the same phenomenology as standard axions and are called axion-like particles (ALPs). The coupling of ALPs to two photons implies the possibility for oscillations from photons into ALPs in an external magnetic field \citep{1983PhRvL..51.1415S,1988PhRvD..37.1237R}. This is used by laboratory experiments to search for ALPs produced in the Sun \citep{2007JCAP...04..010A}, or by conversion of a laser beam in a strong magnetic field \citep{Ehret:2010mh}, to put stringent constraints on the coupling strength. 

Astrophysical environments offer the possibility of strong magnetic fields on long baselines, and thus are promising targets in the search for ALPs. In contrast with laboratory experiments, magnetic fields in astrophysical environments are usually not coherent. In the case of galaxy cluster magnetic fields for instance, the turbulence is well described by a Kolmogorov power spectrum on scales from tens of parsecs up to a few kiloparsecs. It has been shown in \cite{Wouters:2012qd} that when $\gamma$-rays mix with ALPs in a turbulent magnetic fields, the turbulence of the field translates into an irregular behavior of the $\gamma$-ray energy spectrum. These irregularities are expected in a limited energy range, for energies around the energy threshold of the mixing. The search for irregularities in the TeV-energy spectrum of the bright blazar PKS 2155$-$304  measured by H.E.S.S. gives stringent constraints in a limited range for the ALP mass around $10^{-8}\,\mathrm{eV}$ \citep{Wouters:2013moriond}. 

In the present study, X-ray data are used and ALP masses below the effective mass of the photon in the medium are considered. For typical values of the electron density in galaxy clusters, the effective photon mass is about $10^{-11}\,\mathrm{eV}$. Spectral irregularities associated with this low mass are expected in the X-ray band for typical galaxy cluster magnetic fields and values of the coupling $g_{\gamma a}$ that are not excluded by current constraints. One possible source of uncertainty and systematic error in the determination of constraints on ALPs in astrophysics is the lack of knowledge about the magnetic fields that serve as a target to trigger the photon/ALP conversions. Here, we choose to use one of the best characterized large-scale magnetic fields in both strength and structure, that is, the magnetic field of the well-studied Hydra cluster. In the following, irregularities are searched in the spectrum of the bright X-ray source Hydra A that lies at the center of the rich galaxy cluster Hydra. The characteristics of the electron density and those of the magnetic field (profiles, scales, turbulence) in the cluster derived by Faraday rotation measure have been extensively discussed in the literature and are used in the analysis \citep{2003A&A...412..373V,2005A&A...434...67V,2011A&A...529A..13K,2008MNRAS.391..521L}. The article is organized as follows. First, the phenomenology of the \galp oscillations in the medium is briefly recalled. In a second step, the modeling of the magnetic field and electron density in the cluster is presented, following the most recent measurement by Faraday rotation. The results of the analysis of \textit{Chandra} data on the Hydra cluster are then exposed and the derived constraint is finally discussed.

\section{Phenomenology of the photon/ALP system}
\label{sec:pheno}

The photon/ALP system is described as in \cite{1988PhRvD..37.1237R} by a wave function with three states, two for the photon corresponding to the polarization states and one for the axion. The Lagrangian of Equation \ref{eq:lagrangian} induces the mixing between the three states in an external magnetic field. The formalism of the density matrix \citep{2009JCAP...12..004M} is used to compute the probability of observing a photon after propagation in a magnetic domain of size $s$, starting from an unpolarized beam of photons. This probability is called the survival probability. For propagation in one domain with a coherent magnetic field of strength projected on the polarization plane $B$, this probability is written as:
\beq
P_{\gamma \rightarrow \gamma} = 1-\frac{1}{2}\frac{1}{1+E_c^2/E^2}\sin^2\frac{g_{\gamma a}Bs}{2}\sqrt{1+E_c^2/E^2} \;\;,
\label{eq:proba}
\eeq
where $E$ is the energy of the photon and $E_c = |m_\gamma^2 - m_a^2|/2g_{\gamma a}B$ defines the critical energy above which the mixing is efficient. $m_\gamma = 4\pi\alpha n_e/m_e$ is the effective mass of a photon propagating in a plasma with electron density $n_e$. In galaxy clusters, typical values for the electron density are $n_e \sim 0.01 \,\mathrm{cm}^{-3}$, corresponding to an effective photon mass of $m_\gamma \sim 3\times 10^{-11}\, \mathrm{eV}$. For ALP masses $m_a$ negligible compared to $m_\gamma$, the critical energy no longer depends on $m_a$.  The survival probability is energy-dependent for energies around $E_c$. For turbulent magnetic fields, the global transfer matrix of the system has a very complex energy dependence and the survival probability can show strong spectral irregularities around $E_c$. As shown in \cite{Wouters:2012qd}, the exact structure of these irregularities depends on the realization of the turbulent magnetic field crossed by the beam, and therefore is not predictable. However, the statistical properties of the spectral irregularities are a prediction of the model. The following analysis addresses the question of the level of spectral irregularities induced by \galp oscillations that can be accommodated by the data. ALPs with a low mass, lower than the effective photon mass, are considered. In this regime, $m_a \ll m_\gamma \sim 10^{-11}\, \mathrm{eV}$, $E_c$ no longer depends on $m_a$, and the deduced constraint on $g_{\gamma a}$ does not depend on the mass of the ALP. In this mass region, the coupling strength is limited to $g_{\gamma a} \lesssim 10^{-11} \mathrm{GeV}^{-1}$ because of the non-observation of a $\gamma$-ray counterpart to SN 1987A \citep{Brockway:1996yr}. For this upper limit value and typical values of magnetic fields in galaxy cluster, $B \sim 10 \, \mu\rm G$, $E_c$ lies around a few tens of keV, thus motivating observations in X-rays.

\section{Modeling of the galaxy cluster}
\label{sec:model}

For one to obtain constraints on ALP parameters, a bright X-ray point-like source embedded in a strong magnetic field is required. A good knowledge of the magnetic field and electron density is also essential in order to limit possible systematic bias from these uncontrolled parameters. One of the best candidate sources is Hydra A, a Fanaroff-Riley class I radio galaxy located at redshift $z = 0.0538$ that has been observed by the \textit{Chandra Observatory}. It is centered on a corona of hot thermal electrons emitting in X-rays by radiative cooling. The electron density profile from \cite{2005A&A...434...67V} is used. It is estimated with the X-ray surface brightness model of \cite{1999ApJ...517..627M} from \textit{ROSAT} PSPC data, deprojected from the line of sight with the method of \cite{2004A&A...413...17P}. The magnetic field profile surrounding Hydra A has been extensively studied using Faraday rotation maps  of the polarization of the radio emission of the active galactic nucleus lobes due to the propagation through the magnetized electron plasma~\citep{1993ApJ...416..554T,2003A&A...412..373V,2005A&A...434...67V,2008MNRAS.391..521L,2011A&A...529A..13K}. These studies assume that the magnetic field scales with the electron density as $B(r) \propto n_e(r)^{\alpha_B}$, where $\alpha_B$ is a free parameter to be determined. The geometry, described by the angle of projection of the northern jet on the line of sight $\theta$, plays a crucial role in determining the strength of the magnetic field. A depolarization asymmetry is observed between the two lobes of Hydra A \citep{1993ApJ...416..554T}, most likely related to the Laing-Garrington effect \citep{1988Natur.331..147G,1988Natur.331..149L}, because of the non-vanishing $\theta$ angle. \cite{1993ApJ...416..554T} found a most likely value for $\theta$ of $45^\circ$. This value has been confirmed in more recent analyses \citep{2011A&A...529A..13K,2008MNRAS.391..521L} but the associated uncertainty remains large. For instance, a value as low as $30^\circ$ is still plausible \citep{2011A&A...529A..13K}. In the analysis of~\cite{2011A&A...529A..13K}, which is a refinement of \cite{2005A&A...434...67V}, the strength of the magnetic field $B_0$ at the center of Hydra A is found to be 21 $\mu$G if $\theta = 30^\circ$, compared to $B_0 = 36 \, \mu\rm G$ if $\theta = 45^\circ$. To be conservative, the profile of the magnetic field found in \cite{2011A&A...529A..13K} for $\theta = 30^\circ$ is assumed in the following way. The most likely value of the scale parameter is $\alpha_B = 1$, which is significantly higher than the theoretical value of $\alpha_B = 0.5$ that is expected if the ratio of magnetic to kinetic energy is constant throughout the cluster. Again, the value $\alpha_B = 1$ is taken in order to be conservative. Finally, the turbulence power spectrum is found to be in good agreement with a Kolmogorov spectrum $P(k) \propto k^{-5/3}$ up to scales as large as 10 kpc. When it comes to spectral irregularities, it has been shown in \cite{Wouters:2012qd} that \galp mixing rapidly becomes irrelevant for lower scales for two reasons. First, the decreasing Kolmogorov power spectrum suppresses contributions on lower scales as $s^{-2/3}$ \citep{2007PhRvD..76b3001M}. Second, the amplitude of \galp also depends on the turbulence scale, independently of $B$, as $s^{-1/2}$.  For these reasons, in the following, the turbulence power spectrum is modeled with a Kolmogorov slope between scales of 1$-$10 kpc.

\section{\textit{Chandra} data analysis}
\label{sec:analysis}

Hydra A was observed by the ACIS instrument on board the satellite-borne \textit{Chandra Observatory} \citep{2000SPIE.4012....2W} in 1999 and 2003, for, respectively, 20\% and 80\% of the total live time. The ACIS instrument is composed of two arrays of imaging CCDs sensitive to X-rays between 0.2 and 10 keV. It features an average angular resolution of 1$\arcsecond$ and an energy resolution of about 0.1 keV at 1.5 keV \citep{2003SPIE.4851...28G}. The data have been recalibrated using the standard procedure with the latest calibration 4.5.5.1 and the analysis is performed with \texttt{CIAO} tools version 4.5. Events with energy between 0.3 and 10 keV have been retained for the analysis and checked for background event flares. A total live-time exposure of 238 ks is eventually available for the analysis.  
Images of the Hydra A region from \textit{Chandra} have been given in \cite{2000ApJ...534L.135M} and \cite{2009ApJ...707L..69K} and show a diffuse emission from thermal electrons surrounding the central source. In this study, only the non-thermal component of the spectrum from the central source is of interest. To extract this spectrum, an aperture of 1$\arcsecond$ around the position of the central source (determined from infrared observations;~\cite{2006AJ....131.1163S}) is used. Contamination from the thermal background beneath the non-thermal component is estimated from an annular region between 1$\arcsecond$ and 2.5$\arcsecond$. The source spectrum is rebinned to ensure a minimum of 30 counts per bin. The spectral analysis is carried out with the \texttt{XSPEC} package version 12.7.1 and the instrumental response functions generated via the \texttt{specextract} procedure. Spectral fits are performed with a forward folding procedure. The spectrum is well modeled ($\chi^2/n_{\rm d.o.f.}$ = 48.59/58) by a power law ($dN/dE \propto E^{-\Gamma}$, $\Gamma = 1.52 \pm 0.17$) heavily absorbed by a hydrogen column $N_{\mathrm{H}} = 2.54 \pm 0.33 \times10^{22} \, \rm cm^{-2}$ \citep{2000ApJ...534L.135M}. The absorbed integrated flux of the source between 2 and 10 keV is $3.10 \pm 0.07 \times 10^{-5} \, \gamma. \rm cm^{-2} s^{-1}$.
\begin{figure}
\centering
\includegraphics[width =\columnwidth]{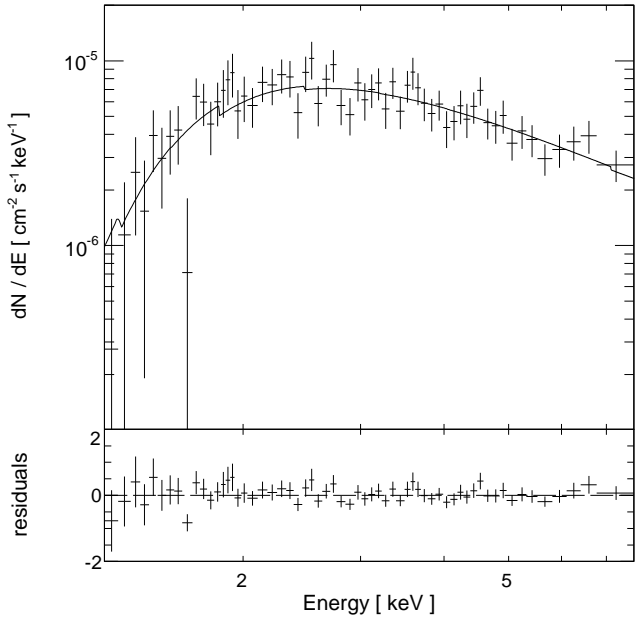}
\caption{Spectrum of the non-thermal component of the point-like source. Top panel: reconstructed spectrum with model. Bottom panel: residuals normalized to the model.
\label{fig:spectrum}}
\end{figure}
The hydrogen column found with the fit is hundreds of times larger than the Milky Way contribution expected in the direction of Hydra A. Evidence for an opaque system at the core of Hydra A has already been reported in \cite{1996ApJ...470..394T}. Due to the heavy absorption at low energies, the non-thermal component from the core of Hydra A is only visible above 1 keV. In the following, the spectrum is restricted to energies above this threshold. The spectrum is shown in Figure ~\ref{fig:spectrum} together with the model. The solid line is the best-fit function, corresponding to an absorbed power law. The small features that appear on the continuous line are related to absorptions due to elements heavier than hydrogen. Also shown are the residuals of the fit normalized to the model. No significant deviations or irregularities that could be linked to \galp oscillations are observed in the spectrum so that it is now used to constrain the value of $g_{\gamma a}$.

\section{The constraint}
\label{sec:constraint}
To estimate the maximum level of irregularities induced by \galp oscillations that can be accommodated by the data, the spectrum is fitted with the model from Section \ref{sec:analysis}, multiplied by an irregularity pattern corresponding to \galp conversions. An example of such a pattern is given in \cite{Wouters:2012qd}. Contrary to \cite{Wouters:2012qd}, the initial photon beam is here assumed to be unpolarized, implying that the survival probability cannot be lower than 0.5. The value of $g_{\gamma a}$ is a new free parameter that drives the amplitudes of the irregularities. When $g_{\gamma a}$ goes to zero, the spectral model without the ALP pattern as shown in Figure ~\ref{fig:spectrum} is retrieved. The exclusion on the additional parameter $g_{\gamma a}$ is obtained using a likelihood estimator $\mathcal{L}$ that assumes that the number of counts in each reconstructed energy bin follows a Poisson statistic.
\begin{figure}[t]
\centering
\includegraphics[width = \columnwidth]{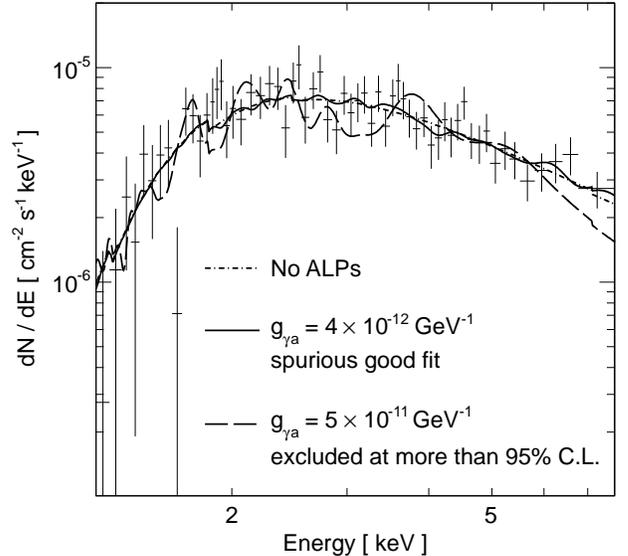}
\caption{Different fits to the data with ALP induced irregularities. Three examples show different signal magnitudes, $g_{\gamma a}=0$ (dot-dashed line), $g_{\gamma a}=4\times 10^{-12}\;\rm GeV^{-1}$ (solid line) and $g_{\gamma a}=5\times 10^{-11}\;\rm GeV^{-1}$ (dashed line).\label{fig:signals}}
\end{figure}
Examples of spectral shapes are displayed in Figure~\ref{fig:signals}, where the model with $g_{\gamma a}=0$ is drawn, along with two spectral shapes corresponding to $g_{\gamma a}=4\times10^{-12}\;\rm GeV^{-1}$ and $g_{\gamma a}=5\times10^{-11}\;\rm GeV^{-1}$.
Due to the turbulent nature of the magnetic field crossed by the beam, the exact structure of the ALP signal is not predictable. The unknown realization of the magnetic field is considered as a nuisance parameter in the analysis. Simulations of the expected signal for a large number of realizations of the magnetic field therefore need to be performed. Note that this randomness accounts for the lack of knowledge on the magnetic field and not its dynamics, with its configuration appearing static over periods of the order of thousands of years for the smallest considered scales. The method used to derive the constraint is based on a likelihood ratio test with nuisance parameters \citep{2005NIMPA.551..493R}:
\beq
\lambda(g_{\gamma a}) =  \frac{\sup\limits_{\theta} \mathcal{L}(g_{\gamma a}, \theta)}{\sup\limits_{g_{\gamma a}, \theta}\mathcal{L}(g_{\gamma a}, \theta)} \;\;,
\eeq
where $\theta$ stands for the nuisance parameters that describe the realization of the magnetic field. 

\begin{figure}[b]
\centering
\includegraphics[width = \columnwidth]{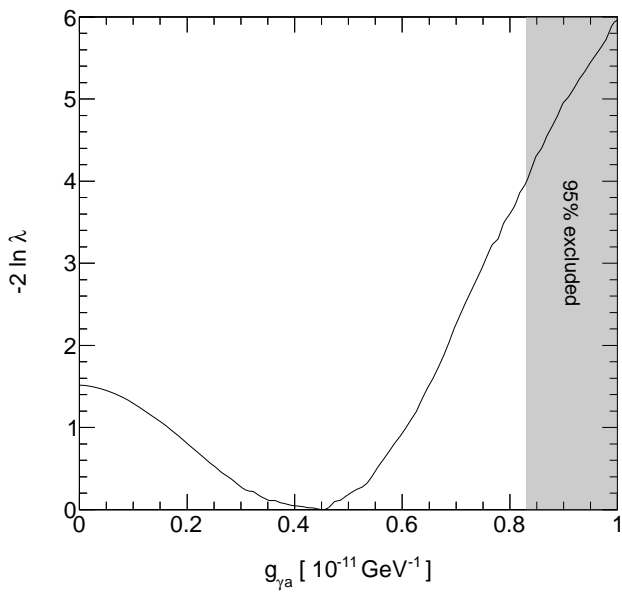}
\caption{Profile of the likelihood  ratio test $-2\ln\lambda$ as a function of $g_{\gamma a}$. The gray filled area gives the exclusion on $g_{\gamma a}$ that is obtained.
\label{fig:profile}}
\end{figure}

The profile of the likelihood ratio test $-2\ln\lambda$ is shown in Figure \ref{fig:profile}.
The best fit is obtained around  $g_{\gamma a} \sim 0.4 \times 10^{-11} \rm GeV^{-1}$, corresponding to a better fit with the ALP pattern compared to the simple absorbed power law. The significance of this maximal likelihood is, however, low at the 1.2$\sigma$ level. This effect is not an indication of a signal but rather an expected effect. For the corresponding realizations of the magnetic field, the spectral irregularities that are produced overfit the Poissonian noise of the data that comes from the finite statistic. At the minimum, the irregularities in the flux have the same amplitude as the shot noise in the bins. One can roughly estimate the value of the coupling constant for which this is expected. Better fits should indeed be found when the scale of the induced irregularities matches the natural statistical fluctuations. There are 30 events per bin in the spectrum of Figure ~\ref{fig:spectrum} so that the relative level of natural irregularity is $1/\sqrt{30}\simeq 20\%$. For photon/ALP induced irregularities, the relative $\delta \phi/\phi$ fluctuations are computed from the average of the survival probability over many domains \citep{2002PhLB..543...23G}:
\beq
\frac{\delta\phi}{\phi} = \frac{1}{3}\left(1-e^{-3g_{\gamma a}^2B^2Ls/8}\right)
\label{eq:average}
\eeq
where $B^2L= \int_0^\infty B^2(r)\mathrm{d}r = 6.5\times10^{-13} \, \rm GeV^3$ is the integrated magnetic field profile over the line of sight and $s = 10$ kpc is the coherence length. The irregularities are then comparable to the statistical fluctuations for $g_{\gamma a}$ of a few $10^{-12} \,\rm GeV^{-1}$, in agreement with the minimum found in Figure~\ref{fig:profile}.  The value of $g_{\gamma a}$ at this minimum is thus predicted as a function of the magnetic field parameters and the statistic in the bins. For this reason, the minimum does not correspond to an indication for an ALP. Moreover, there is no guarantee on the exact spectral model underlying the data. For instance, there could well be line features due to the fluorescence of iron or other heavy elements that would mimic the ALP pattern.  In the end, this method could not be used for a discovery, but gives conservative constraints. For higher coupling strengths, the irregularities become larger than the typical range of noise induced by the Poisson statistic and the fit degrades quickly. As an illustration, Figure~\ref{fig:signals} exhibits two spectral shapes where the ALP signal is present. In Figure~\ref{fig:signals}, the dot-dashed line is the conventional fit, the solid line is an example of a spurious good fit, that is obtained when the ALP induced fluctuations are of the same magnitude as the natural fluctuations. The dashed line corresponds to a large value of $g_{\gamma a}$, that induces large fluctuations that are at odds with the measurements.

\begin{figure}[t]
\centering
\includegraphics[width = \columnwidth]{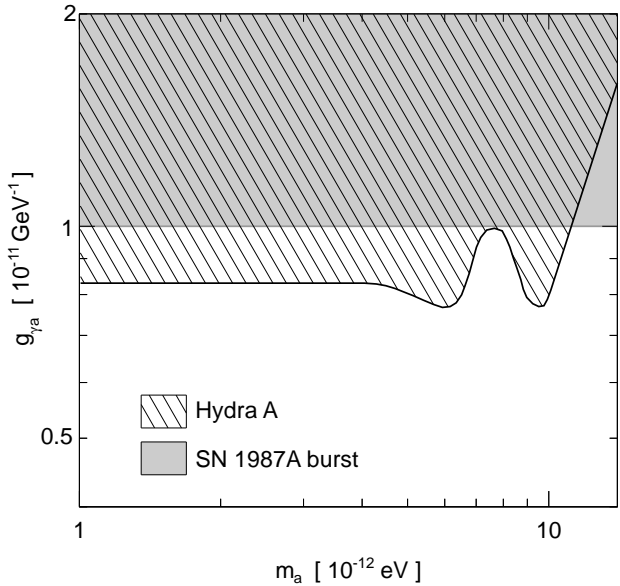}
\caption{Constraint on $g_{\gamma a}$ from X-ray observations of Hydra A cluster (hatched area). The grey filled area is the constraint from SN 1987 A burst~\citep{Brockway:1996yr}. Exclusions are set at the 95\% C.L.
\label{fig:constraint}}
\end{figure}

The test $-2 \ln \lambda$ follows a $\chi^2$ distribution with one degrees of freedom \citep{2005NIMPA.551..493R} so that values of $g_{\gamma a}$ for which the test is larger than 4 are excluded at the 95\% confidence level (CL).
This yields the constraint $ g_{\gamma a} < 8.3 \times 10^{-12} \,\rm GeV^{-1}$ at the 95\% C.L. This constraint is derived for ALP models with arbitrarily small masses $m_a$. When the ALP mass compares to the effective photon mass in the plasma, the spectral irregularities are no longer independent of $m_a$. As seen in Section~\ref{sec:pheno}, this occurs when $m_a \sim 10^{-11} \,\rm GeV^{-1}$. The procedure described above is thus repeated for different values of $m_a$ in order to derive the shape of the limit in this range of mass. The contour of exclusion obtained at the 95\% CL is shown in Figure~\ref{fig:constraint}. There is a small range of mass around $8\times 10^{-12} $ eV where the exclusion is slightly less constraining. This corresponds to ALP masses comparable to the effective photon mass in the plasma. From Equation~\ref{eq:proba}, the survival probability is energy independent when $m_a = m_\gamma$. In this case, no spectral irregularities occur and no constraint can be given. However, when considering the propagation in the Hydra A cluster, the electron density is not uniform and decreases along the beam path. At higher ALP masses, the exclusion becomes less constraining as the critical energy increases to energies that cannot be probed with $\textit{Chandra}$. Therefore, the exclusion curve rises as $m_a^2$ as expected.
\cite{Horns:2012pp} and \cite{2012PhRvD..86k5025T} also set a constraint on very low mass ALPs, but due to the lack of knowledge on the magnetic field properties used for their analysis, the authors only constrain the $g_{\gamma a}B$ product. The constraint obtained from the burst of SN 1987 A in~\cite{Brockway:1996yr} is also shown in Figure~\ref{fig:constraint}. For these small masses, direct searches at CAST yield a limit of the order of $10^{-10}\;\rm GeV^{-1}$ \citep{2007JCAP...04..010A}. It appears that the present constraint is therefore the most competitive for low-mass ALPs. Future laboratory experiments such as ALPS-II~\citep{Bahre:2013ywa} and IAXO~\citep{2013arXiv1302.3273V} will improve the constraints in this region of the parameter space.

A final remark is that the current limits hold for a more general class of particles than ALPs. Indeed they are valid for generic pseudo-Nambu-Goldstone bosons, even in the case of an $F^2$ type of coupling to photons instead of the $F\tilde{F}$ term introduced in Equation~\ref{eq:lagrangian}. Switching to the first type of coupling leads to a swap of the polarization of the photon that is involved in the mixing. As magnetic field directions are randomly chosen inside many domains here, the effect of the mixing on the irregularities is the same.

\section{Discussion and conclusion}
\label{sec:discussion}
The search for irregularities in the X-ray spectrum of Hydra A measured by \textit{Chandra} leads to a constraint at the 95\% CL on the ALP coupling to two photons of $g_{\gamma a} < 8.3 \times 10^{-12}\, \mathrm{GeV}^{-1}$ for ALP masses below $7\times 10^{-12}\,\mathrm{eV}$. In this study, the key point is the use of a measured profile of the magnetic field, determined from Faraday rotation maps.  The related measurement of the magnetic field yields a determination of its strength that has an error, the lower limit of which is retained in the present analysis to be conservative. The constraints are thus considered as firm and robust. For a more likely profile of the magnetic field, the constraint would improve to $g_{\gamma a} < 4.3 \times 10^{-12}\, \mathrm{GeV}^{-1}$. A refined measurement of the jet orientation angle $\theta$ of Hydra A would help to characterize the magnetic field profile and possibly improve the limit. The constraint derived here is limited by the available statistics. Simulations of spectra that would be observed with more data show that with a 10 times longer exposure, the upper limit could be improved to $g_{\gamma a} < 3.9 \times 10^{-12}\, \mathrm{GeV}^{-1}$ for the conservative magnetic field profile. The conservative exclusion derived in this work is the most stringent constraint to date in the range of very low mass ALPs, with $m_a < 10^{-11}\,\mathrm{eV}$.

\acknowledgments

This work benefited from the support of the French PNHE (Programme National Hautes Energies).



\end{document}